\documentclass[amsmath,amssymb,superscriptaddress,aps,prl,twocolumn,reprint]{revtex4-1}

\usepackage{graphicx}

\usepackage{amssymb,amsfonts,amsmath}
\usepackage{bm}
\usepackage{color}

\renewcommand {\vec} [1] {{\bm #1}}
\newcommand{\ff}{g}
\newcommand{\tf}{h}
\newcommand{\mat}[1]{#1}
\newcommand{\noise}{\eta}

\begin{document}


\title{Compressed sensing for multidimensional electronic spectroscopy experiments}


\author{J. N. Sanders$^1$, S. Mostame$^1$, S. K. Saikin$^1$, X. Andrade$^1$, J. R.  Widom$^2$, A. H. Marcus$^2$, and A. Aspuru-Guzik$^*$}
\address{Department of Chemistry and Chemical Biology, Harvard University, 12 Oxford Street, Cambridge, MA 02138, United States
  \\
$^2$Department of Chemistry, Oregon Center for Optics, Institute of Molecular Biology, University of Oregon, Eugene, OR 97403, United States
}



\begin{abstract}
Compressed sensing is a processing method that significantly reduces the number of measurements needed to accurately resolve signals in many fields of science and engineering.  We develop a two-dimensional (2D) variant of compressed sensing for multidimensional electronic spectroscopy and apply it to experimental data.  For the model system of atomic rubidium vapor, we find that compressed sensing provides significantly better resolution of 2D spectra than a conventional discrete Fourier transform from the same experimental data.  We believe that by combining powerful resolution with ease of use, compressed sensing can be a powerful tool for the analysis and interpretation of ultrafast spectroscopy data.
\end{abstract}

\maketitle






Multidimensional spectroscopy
\cite{Hochstrasser_PNAS2007,Hamm2001,Khalil2003,Jonas2003} is an
important tool for studying ultrafast dynamical processes in
complex molecular systems. For instance, it can be used to analyze
vibrational energy transfer at liquid/air interfaces on picosecond
timescales~\cite{WaterNatChem2011} or exciton dynamics in natural
light harvesting systems at hundreds of
femtoseconds~\cite{Brixner_Nat2005, YuenZhou2011, YuenZhou2011-2}.
It can also be applied for efficient detection and identification
of molecules, which is one of the crucial challenges in
chemistry, biology, and medicine with important applications, for
example, in molecular sensing, chemical separation, and DNA
analysis.
Frequently, in these nonlinear optical techniques, the data
collected in the time domain is Fourier transformed to the
frequency domain.  The crucial issue, then, for obtaining high frequency resolution in measured spectra is the long sampling time required.
Here, we demonstrate that a method known as compressed sensing (CS) can be applied
as a very efficient alternative to the Fourier transform to obtain
high-resolution multidimensional spectra.

Compressed sensing is a state-of-the-art signal processing method which has recently
become popular throughout the physical and biological sciences.
The method is founded on the concept of \emph{sparsity}.  When a signal is known to be sparse
in a certain basis (i.e. most of the coefficients are negligibly small), this additional
knowledge can be used to dramatically reduce the number of
measurements required to reconstruct the signal~\cite{Candes2006,Donoho2006}.
This method has been applied to many areas of
research, ranging from magnetic resonance
imaging~\cite{Lustig2007} to superresolved imaging of single molecules
~\cite{Zhu2012} and quantum process
tomography~\cite{Shabani2011}.
Earlier, some of us showed that CS can also be used to
significantly reduce the computational cost of atomistic
simulations~\cite{Andrade2012-2}. In that work, the application of
CS was particularized for molecular dynamics and real-time
time-dependent density functional theory
simulations for obtaining linear spectra (vibrational, optical absorption, and circular dichroism)~\cite{Andrade2012}.  However, CS can be applied to
other types of simulations and experimental techniques.
Multidimensional non-linear spectroscopy is then an interesting and
relevant candidate to explore the possibilities of CS.
Recently, CS has been pursued to reconstruct one of the dimensions in ultrafast 2D NMR data~\cite{Shrot2011}
and to simulate 2D spectra using random sampling~\cite{Almeida2012}.

In this letter, we present the first application of CS to
experimental ultrafast 2D optical spectroscopy.
We apply the method to an atomic system -- Rubidium vapor, which
is frequently used as a test model for multidimensional
spectroscopy techniques~\cite{Warren2003,Vaughan2007,Andy2007}.
We find that CS presents several additional advantages
beyond a simple speed-up.  Many multidimensional experiments are inherently limited in the amount
of time-domain data that may be collected, either due to measurement constraints or to more fundamental
limitations such as the timescale of the dynamics one wishes to explore (which may be very short due
to decoherence and other processes).  We believe CS can extend the range of spectroscopically-observable
dynamics by providing higher frequency resolution even given limited time domain data.

Another benefit of CS is that it is quite easy for experimentalists to integrate.
As discussed below, our approach simply replaces the 2D discrete Fourier transform
(FT) by a new 2D compressed sensing scheme.  Since we recast 2D CS as a series of
one-dimensional problems, we are able to harness the power of parallel computing so
that the entire signal processing does not take significantly longer than 2D FT.
It is our hope that CS's easy portability and short processing time will help it become a method of
choice among experimental spectroscopists.

The rest of this letter is structured as follows.  We begin by presenting the compressed sensing method and
outlining its application to the resolution of 2D ultrafast optical spectra.  We next apply the
method to a model experimental system, namely gas-phase Rubidium atoms, and show how CS may be used
to better resolve the spectral features.  We then discuss the experimental methods and numerical implementation.
Finally, we offer conclusions and a future outlook.
\begin{figure}
  \centering
  \includegraphics*[width=\columnwidth]{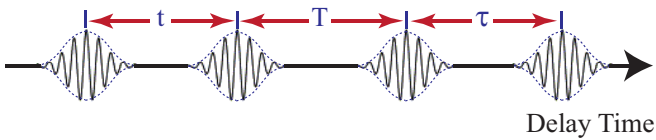}
  \caption{Schematic illustration of the sequence of four pulses used in 2D optical spectroscopy.}
  \label{fig:pulses}
\end{figure}
We begin by describing how the CS method 
can be applied to 2D ultrafast optical spectroscopy;
see refs.~\cite{Baraniuk2007,Candes2008,Chartrand2010}  for more detailed information about CS.
In particular, we will focus on four-wave mixing experiments, similar to those used to study coherent energy transfer in light-harvesting complexes, quantum dots, and other systems of physical and biological interest.  Typically, these experiments involve irradiating a sample with four optical pulses and varying the time gaps between the pulses (Fig.~\ref{fig:pulses}).  Defining the time gaps as \(\tau\) (the coherence time between pulses 1 and 2), \(T\) (the population time between pulses 2 and 3), and \(t\) (the waiting time between pulses 3 and 4), the signal measured in the time domain is a function \(S(\tau,T,t)\).  The standard approach in four-wave mixing is to perform a 2D discrete Fourier transform in \(\tau\) and \(t\) to obtain the spectrum \(S(\omega_\tau,T,\omega_t)\).  This spectrum is typically plotted in \(\omega_\tau\)---\(\omega_t\) space for different values of \(T\), and the dynamics of the peaks then give information about the dynamics of the underlying system.

The CS method can be used to replace a 1D discrete FT in converting
time-resolved data into the frequency domain~\cite{Candes2006,Andrade2012-2},
obtaining a significantly better frequency resolution than the
conventional discrete FT.  Our basic approach here, then, is to reduce
the 2D CS computation to a series of 1D CS calculations.  A major
advantage of this approach (as compared to an inherently 2D method) is
that the 1D computations can be performed in parallel, and each is
individually inexpensive.  To obtain the 2D
spectrum for a particular population time \(T\), our procedure
involves two steps:

\emph{Step 1:} For every value of \(\tau\) in the data set, perform a 1D CS in \(t\) to convert the signal to the \(\omega_t\) domain.  By collecting all the results for different \(\tau\) together, we obtain the half-transformed signal \(S(\tau,T,\omega_t)\) which is sparse in \(\omega_t\).  This step requires a total of \(N_\tau\) 1D CS calculations (where \(N_\tau\) is the number of \(\tau\) data points).  Each calculation takes a short time and all of them can be performed in parallel.

\emph{Step 2:} Now for every value of \(\omega_t\) in the half-transformed signal, perform a 1D CS in \(\tau\) to convert the signal to the \(\omega_\tau\) domain. This step requires \(N_{\omega_t}\) 1D CS calculations which also take a short time and may be performed in parallel.  Collecting all the results together yields the final sparse spectrum \(S(\omega_\tau,T,\omega_t)\).

It should be noted that the roles of \(\tau\) and \(t\) may be interchanged to yield similar final results.  In either case, if the spectrum is sparse in the \(\omega_\tau\)---\(\omega_t\) plane, then this 2D compressed sensing procedure yields better-resolved peaks in the frequency domains with less time-domain data than the conventional 2D discrete FT, as we will illustrate.

For completeness, however, we first provide  
a very brief review
of the 1D CS calculations, as these underlie the 2D method discussed
above (see ref.~\cite{Andrade2012-2} for a more detailed derivation).
In each 1D CS calculation, we want to obtain a vector \(\vec{\ff}\)
of values \(\{\ff_1, \ff_2, \ldots, \ff_{N_\omega}\}\) at \(N_\omega\)
equidistant frequencies \(\omega_j=\Delta\omega\,j\), from the known vector \(\vec{\tf}\) set of time-resolved values $\{\tf_1, \tf_2, \ldots, \tf_{N_t}\}$ given at \(N_t\) equidistant times \(t_j=\Delta t j\).  (The scheme can be generalized for non-uniform sampling.)  Our objective is to obtain sensible results with \(N_t\) as small as possible, so we are interested in the case \(N_\omega > N_t\).
In principle, the \(\ff_k\) set can be directly obtained using the
discrete FT,
\begin{equation}
  \label{eq:ft}
  \ff_k = \sum_{j=1}^{N_t} \Delta t \, e^{i \omega_k t_j}\,\tf_j\ .
\end{equation}
However, if we expect that many of the Fourier coefficients are
negligible, we can use CS to
extract more precise results from the same signal. This is done by the
recasting the Fourier coefficient calculation as a linear
equation. Since this is an underdetermined problem (as \(N_\omega >
N_t\)), the sparsity condition means that we should select the solution
that has the larger number of zero coefficients. In practice this
solution can be obtained by solving the basis-pursuit de-noising (BPDN) problem~\cite{Candes2006}
\begin{equation}
  \label{eq:bpdn}
  \min_{\vec{g}} |\vec{\ff}|_1 \quad \textrm{subject to}\quad
  \left|\mat{F}\vec{\ff}-\vec{\tf}\right|_2 < \noise\ ,
\end{equation}
where \(\mat{F}\) is the \(N_\omega\times N_t\) inverse Fourier matrix with entries
\begin{equation}
  \label{eq:f}
  F_{jk} = \frac{2}{\pi}\Delta\omega \, e^{-i \omega_j t_k}
\end{equation}
and \(\noise\) represents a level of noise that we assume is present in the signal.  In all calculations which follow, \(\noise < 10^{-4}\).

\begin{figure}
  \centering
  \includegraphics*[width=0.7\columnwidth]{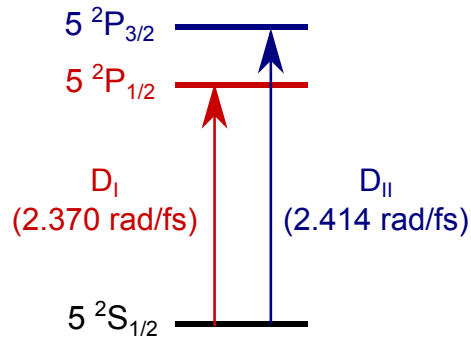}
  \caption{Energy level diagram for atomic \(^{87}\)Rb vapor.}
  \label{fig:rb}
\end{figure}

To illustrate the utility of CS in two-dimensional optical spectroscopy, we consider phase-modulation 2D fluorescence spectroscopy (PM-2DFS) data collected from atomic \(^{87}\)Rb vapor~\cite{Andy2007}.  The \(^{87}\)Rb system may be considered as a quantum three-level system with ground state \(5\ ^2S_{1/2}\), first excited state \(5\ ^2P_{1/2}\), and second excited state \(5\ ^2P_{3/2}\), as illustrated in Fig.~\ref{fig:rb}.  The four light pulses are produced by a titanium saphire laser [with full width at half maximum (FWHM) \(\sim\) 42~fs] which is on resonance with the two electronic D line transitions of \(^{87}\)Rb: \(5\ ^2S_{1/2} \rightarrow 5\ ^2P_{1/2}\) (with transition frequency 2.370~rad/fs) and \(5\ ^2S_{1/2} \rightarrow 5\ ^2P_{3/2}\) (with transition frequency 2.414~rad/fs).  We considered \(^{87}\)Rb vapor to be an ideal candidate for our 2D CS method as its 2D electronic spectrum is expected to be sparse in frequency space, with diagonal and cross peaks corresponding to the transitions just mentioned.  The natural lifetime of electronic excitations in \(^{87}\)Rb is \(\sim25 \textrm{ ns}\), which should correspond to very narrow lines (D\(_I\) 36.1 and D\(_{II}\) 38.1 rad/ms).
Full and extensive details of the PM-2DFS experimental method used to collect the data are given in refs.~\cite{Andy2007,Lott2011}.  The experiments yield time-resolved fluorescence-detected ``sum'' and ``difference'' signals [\(S_{\textrm{sum}}(\tau,T,t)\) and \(S_{\textrm{diff}}(\tau,T,t)\)] which are analogous, respectively, to the nonrephasing and rephasing third-order polarizations collected in more traditional four-wave mixing experiments, minus the effects of nonresonant interactions.

\begin{figure*}
\begin{center}
  \includegraphics*[width=\textwidth]{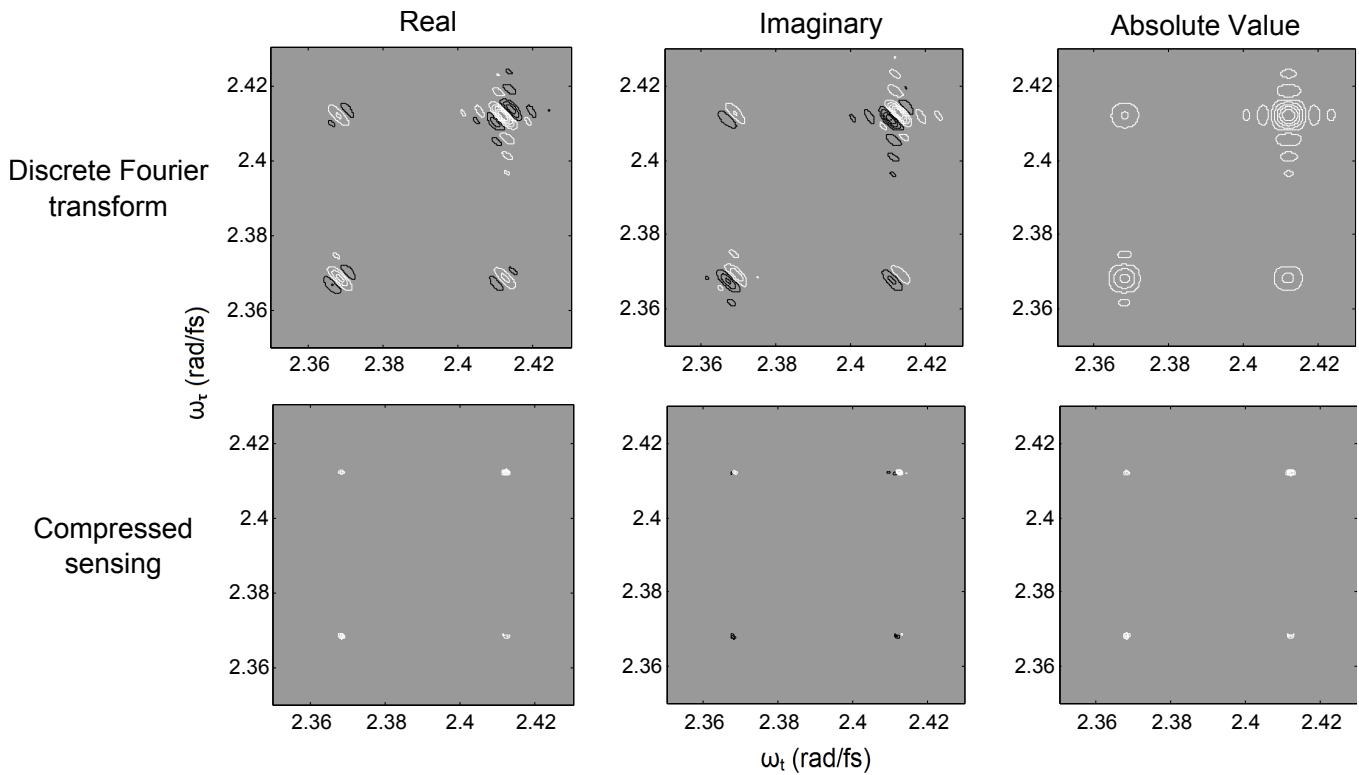}
  \caption{Comparison between discrete Fourier transform (top) and compressed sensing (bottom) for the ``sum'' (nonrephasing) 2D optical spectra of \(^{87}\)Rb vapor for population time \(T = 140 \textrm{ fs}\).  CS yields narrower, better-resolved peaks than the discrete FT.}
  \label{fig:sum}
\end{center}
\end{figure*}

\begin{figure*}
  \centering
  \includegraphics*[width=\textwidth]{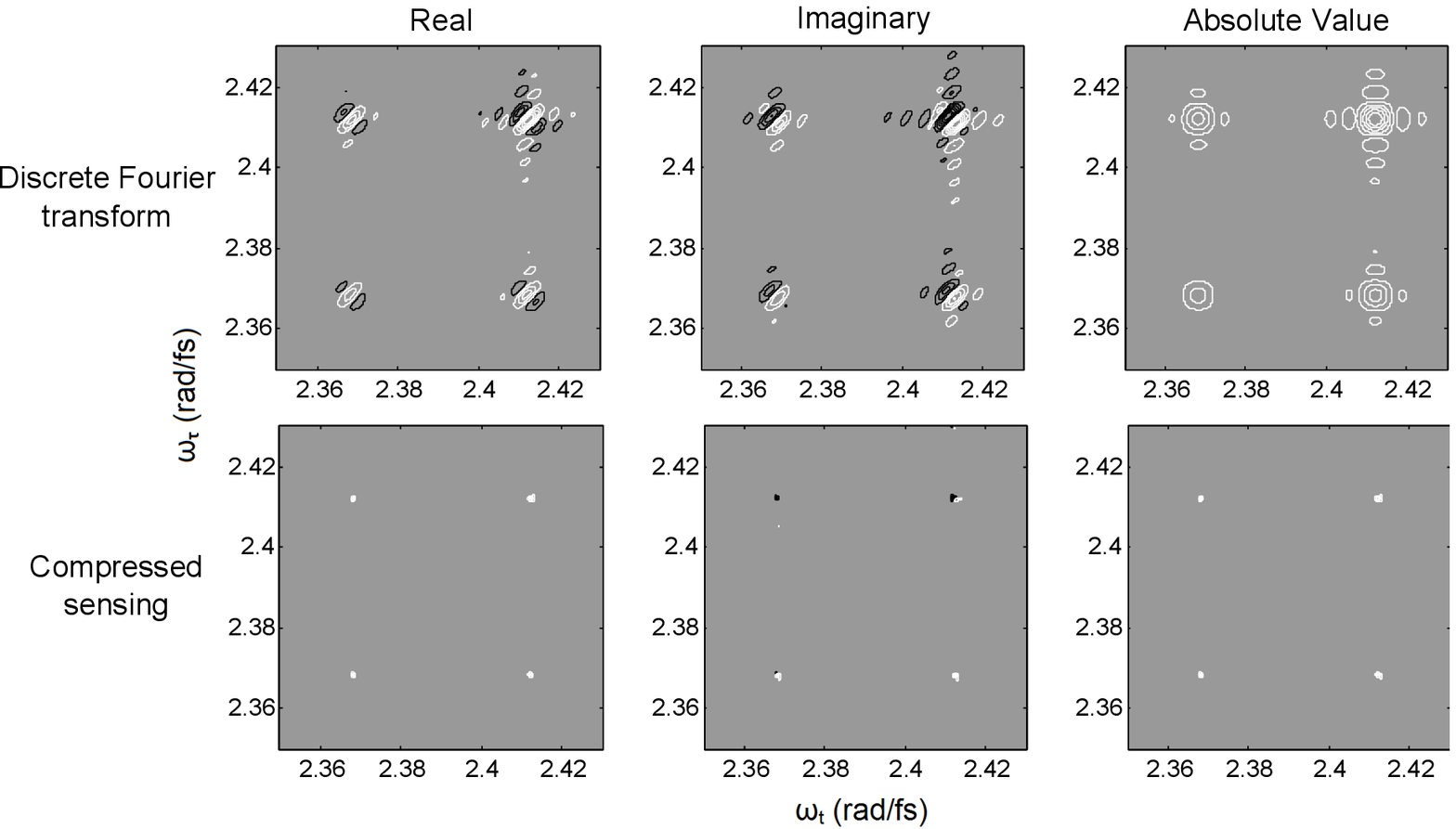}
  \caption{Comparison between discrete Fourier transform (top) and compressed sensing (bottom) for the ``difference'' (rephasing) 2D optical spectra of \(^{87}\)Rb vapor for population time \(T = 140 \, \textrm{ fs}\).  CS yields narrower, better-resolved peaks than the discrete FT.}
  \label{fig:diff}
\end{figure*}

The main results of this letter are given in Figs.~\ref{fig:sum} and \ref{fig:diff}, which compare the performance of the conventional discrete Fourier transform (top) and our 2D compressed sensing method (bottom) in resolving the peaks corresponding to transitions in the atomic \(^{87}\)Rb vapor.  Fig.~\ref{fig:sum} shows the ``sum'' (nonrephasing) signal \(S_{\textrm{sum}}(\omega_\tau,T,\omega_t)\) while Fig.~\ref{fig:diff} shows the ``difference'' (rephasing) signal \(S_{\textrm{diff}}(\omega_\tau,T,\omega_t)\) for the population time \(T = 140\,  \textrm{fs}\).  Similar results were obtained for other population times. Note that the exact same set of time-resolved data was used for the discrete FT spectra as for the CS spectra.

As can be seen from the figures, CS produces peaks that are far better resolved in frequency space than those obtained by the discrete FT.  In fact, the peaks obtained by CS are at least 5 to 10 times less wide in each dimension than those obtained via the discrete FT, consistent with the results we previously obtained in one-dimensional spectra~\cite{Andrade2012-2}.
In other words, to achieve a comparable resolution as CS with the discrete FT, one would need to collect far more time-resolved data by continuing to measure the signal \(S(\tau, T, t)\) for longer time delays \(\tau\) and \(t\).  Given that CS obtains narrow peaks with only limited time-resolved data, we expect this signal processing method
to be particularly useful in resolving closely-spaced peaks in their 2D spectra.

One immediate question arising from the comparison of the FT and CS spectra in Figs.~\ref{fig:sum} and \ref{fig:diff} is that of peak shape.
As for long sampling times the CS result matches the FT spectra, the question
is how capable is CS of resolving those peak features for short times.
The issue of peak shape certainly deserves a more extensive
investigation in the future, most likely by applying our 2D CS method
to more complex experimental systems with intricate features that are
broadened by internal structure or an environment.

In conclusion, we have demonstrated the first application of compressed sensing to two-dimensional electronic spectroscopy experiments.  Focusing on electronic transitions in an atomic \(^{87}\)Rb vapor model system, we have shown that the 2D CS method that we have developed provides much finer resolution of the peaks in a 2D spectrum as compared to the standard discrete FT.  As a result, we expect that 2D CS will substantially reduce the experimental effort needed to obtain well-resolved spectra by decreasing the amount of time-resolved data which must be collected.  Furthermore, we hope that 2D CS will increase the range of spectral features resolvable in ultrafast experiments, particularly closely-spaced peaks.

An important question is the degree to which CS accurately reproduces peak shapes in more complex systems with internal structure and an environment; this question definitely merits a further joint theoretical and experimental effort.  We hope and expect that CS will become more widely investigated and employed in the 2D experimental ultrafast community once its easy portability, competitive speed, and strong resolving power become widely known.

\section{Methods}

The experimental methods used to collect the time-resolved data for the atomic \(^{87}\)Rb vapor are detailed in ref.~\cite{Andy2007}.  Time-resolved ``sum'' and ``difference'' signals were collected at all points on a two-dimensional grid consisting of 51 equally-spaced coherence times \(\tau\) and 50 equally-spaced waiting times \(t\) (with \mbox{\(\Delta\tau = \Delta t = 26.687 \textrm{ fs}\)}) for a series of 5 population times \(T = 140, 175, 210, 245, 280 \textrm{ fs}\).

The full 2D CS spectrum was obtained by performing a series of 1D CS calculations as described in the text.  For each 1D CS calculation, the optimization problem in eq.~(\ref{eq:bpdn}) was solved using the SPGL1 algorithm developed by van~den~Berg and Friedlander~\cite{Berg2008}.  To avoid numerical stability issues we work with a normalized BPDN problem, where the prefactor \(2\Delta\omega/\pi\) of the \(F\) matrix, eq.~(\ref{eq:f}), is left out and \(\vec{\tf}\) is normalized. The missing factors are included in \(\vec{\ff}\) after the solution is found.

For the CS calculations, we use a frequency grid consisting of 1000 evenly-spaced points between \(-\pi/\Delta \tau\) and \(\pi/\Delta \tau\) for \(\omega_\tau\) and 1000 evenly-spaced points between \(-\pi/\Delta t\) and \(\pi/\Delta t\) for \(\omega_t\) (the same grid spacings are also used for the discrete FT).





\begin{acknowledgments}
We acknowledge J.~Yuen-Zhou and J.~Goodknight for useful discussions. 

The computations in this paper were run on the Odyssey cluster supported by the FAS Science Division Research Computing Group at Harvard University.

This work was supported by the Defense Threat Reduction
Agency under Contract No HDTRA1-10-1-0046 and by the Defense Advanced Research Projects Agency under award number N66001-10-1-4060.  J.N.S. acknowledges support from the Department of Defense (DoD) through the National Defense Science \& Engineering Graduate Fellowship (NDSEG) Program.  Further, A.A.-G. is grateful for the support of the Camille and Henry Dreyfus Foundation and the Alfred P. Sloan Foundation.

\end{acknowledgments}

$^*$\,{\footnotesize\sf aspuru@chemistry.harvard.edu}






%
\bibliography{biblio}









\end{document}